\documentclass[proof]{WileyASNA-v2}

\articletype{Original Article}

\received{X X 2023}
\revised{X X 2023}
\accepted{X X 2023}

\usepackage[T1]{fontenc}
\usepackage[utf8]{inputenc}
\usepackage[american]{babel}

\begin{document}

\title{Bright Spectroscopic Binaries: III. Binary systems with orbital periods of $P>500$~days}

\author[1]{Dennis Jack*}
\author[1]{Missael Alejandro Hern\'andez Huerta}
\author[1]{Faiber Danilo Rosas-Portilla}
\author[1,2]{Klaus-Peter Schr\"oder}

\authormark{D. Jack et al.}

\address[1]{\orgdiv{Departamento de Astronom\'\i{}a}, \orgname{Universidad de Guanajuato}, \orgaddress{\state{Guanajuato}, \country{Mexico}}}
\address[2]{\orgdiv{Sterrewacht Leiden}, \orgname{Universiteit Leiden}, \orgaddress{\state{Leiden}, \country{Netherlands}}}

\corres{*Dennis Jack, \email{dennis.jack@ugto.mx}}

\presentaddress{Departamento de Astronom\'\i{}a, Universidad de Guanajuato, Callej\'on de Jalisco S/N, 36023 Guanajuato, GTO, Mexico}

\abstract{We present an analysis of nine bright spectroscopic binaries 
(HD~1585, HD~6613, HD~12390, HD~39923, HD~55201, HD~147430, HD~195543, HD~202699, HD~221643),
which have orbital periods of $P > 500$~days. These well-separated binaries
are the last stars of our sample that we observed with
the TIGRE telescope obtaining intermediate-resolution spectra of $R\approx 20,000$.
We applied the same method as described in our previous publication of this series.
For the analysis of the radial velocity curves, we used the toolkit RadVel,
which allowed us to determine all orbital parameters.
Recently published orbital solutions of some systems from Gaia~DR3 agree with our results.
However, our solutions have much smaller uncertainties. 
We determined the basic stellar parameters of the primary stars with our automatic script using iSpec.
The parameter determination allowed us to place all nine stars in the Hertzsprung-Russell diagram.
We found that all stars have already evolved to the giant phase.
A comparison with stellar evolution tracks of the Eggleton code was applied to
determine the stellar masses and ages. 
As a result of our analysis, we were able to estimate the masses of the secondary stars and the
orbital inclinations of the binary systems.
}


\fundingInfo{UG CIIC 2023}

\maketitle

\footnotetext{\textbf{Abbreviations:} RV, radial velocity;}

%
% Introduction
%
\section{Introduction}

Our Galaxy, the Milky Way, contains a vast amount of double or multiple stellar systems.
Well-separated binaries like in this paper form a large fraction 
of the stellar galactic content \citep{fisher05}.
In the solar neighborhood a fraction for all type of binaries of about 50\% has been estimated \citep{raghavan10}.
Furthermore, it has been shown that the formation of multiple stellar systems 
is actually a natural outcome of the star formation process \citep{tohline02,binreview}.
Binary stars are still, if not the only anymore, but the best source
of stellar mass information, which we need to test our evolution models in a quantitative way.
In addition, the age of both components is the same, which sets precise
constraints to any analysis. Especially of the rarer type of binaries
with a giant, it is therefore in the interest of stellar astrophysics to know more systems.

Recently, the Gaia mission \citep{gaiamission} published 
its third data release (Gaia DR3)
based on about 34 months of observations and data collection \citep{gaiadr3}.
The Gaia DR3 contains now a non-single stars catalog \citep{gaianonsingle}, which
contains about 813,000 objects discovered and analyzed with different detection methods. 
Because of
the nature of the Gaia mission and telescope most of the binaries were observed 
with astrometric methods, which include also a proper motion analysis.
However, the catalog contains also eclipsing binaries and spectroscopic binaries.
Spectroscopic binaries are detected by periodic variations in the radial velocity (RV).
Usually, only the primary star is visible in the spectrum (SB1), but if the two stars have similar
luminosities, spectral lines of the secondary star will appear in the observed spectrum (SB2).
The non-single star catalog of Gaia~DR3 contains also the orbital parameters of some SB2 stars.
There are also orbital solutions in the catalog where different methods have been combined,
e.g. astrometry and spectroscopy.

In our previous publications of this series about bright spectroscopic binaries 
(\citet{paper1} (hereafter Paper~I) and \citet{paper2}(hereafter Paper~II)),
we reported the detection of 19 bright spectroscopic binaries ($m_V < 7.66$~mag) 
and presented the detailed analysis
of ten of these binaries that have periods of $P<500$~days. 
In this third publication of the series, we present a detailed study of the nine remaining 
bright spectroscopic binaries with periods of $P>500$~days. 
Such large periods allow the assumption, that the two components
evolved separately, i.e., without mass exchange (see also \citet{fisher05}). Hence, we
here apply the well-tested single star Eggleton-code evolution models \citep{pols97,schroder97}.
Cases with known sin i allow then to further infer the secondary mass
and provide tests to the models.

Please consult also the first two papers for a more detailed introduction and motivation.
The method used in this paper is basically the same as described in Paper~II. Some of the stars have
already orbital solutions in the non-single stars catalog of Gaia~DR3, and we will compare
our results to these solutions.

%
% Observation and orbital parameters
%
\section{Binary Orbits}\label{sec:orb}

\subsection{Observations}

For the observations of the time series of spectra we used the 1.2~m robotic telescope TIGRE 
\citep{schmitt14,tigre22}.
The spectrograph has an intermediate resolution of $R\approx 20,000$ in a spectral range from about $3800$ to 
$8800$~\AA, with a small gap of about 120~\AA\ around 5800~\AA.
Because of the brightness of the stars, we obtained each individual spectrum with an exposure
time of just one minute.
For the automatic determination of the RVs,
we used the same procedure as described in Papers~I and II, and in \citet{mittag18}.
First, telluric lines are used to better calibrate the wavelength. 
Then, a cross-correlation with a reference spectrum
taken from a set of synthetic spectra using PHOENIX models \citep{Husser2013}
is used to determine the RV. This procedure has mean precision of
about $0.11$~km~s$^{-1}$ \citep{mittag18}.

We made all individual RV measurements electronically available as supporting
online material at \emph{link to Wiley page here} and at 
\url{http://www.astro.ugto.mx/~dennis/binaries/}.

\subsection{Orbital parameters}

To determine the orbital parameters of all nine spectroscopic binaries,
we used the Radial Velocity Modeling Toolkit RadVel \citep{radvel} (version 1.4.9).
The toolkit was designed for exoplanets but worked also very well with stellar binaries.
The orbital parameters were the orbital period $P$, the time of inferior conjunction $T_c$,
eccentricity $e$, semi-amplitude $K$, the argument of the periapsis
of the star's orbit $\omega$ and the RV of the system $v_\mathrm{rad}$.
RadVel fitted the observed RV curves by solving the set of equations for the Keplerian orbit by
an iterative method.
We used the Markov-Chain Monte Carlo (MCMC)
package of \citet{foreman2013} included in RadVel to obtain the uncertainties
for our orbital solution.

\begin{table*}[t]
\centering
\caption{The orbital parameters of nine spectroscopic binary systems determined with the toolkit RadVel.}
\tabcolsep=0pt%
\begin{tabular*}{500pt}{@{\extracolsep\fill}lcccccc@{\extracolsep\fill}}
\toprule
\textbf{Star} & $P$ (days) & $T_c$ (JD) & $e$ & $K$ (km~s$^{-1}$) & $\omega$ (rad) & $v_\mathrm{rad}$ (km~s$^{-1}$) \\ 
\midrule
HD 1585   & $1085.8 \pm 4.4$  & $8203.6 \pm 5.3$  & $0.095 \pm 0.0088$  & $7.494  \pm 0.056$ & $1.699 \pm 0.088$ & $35.841 \pm 0.042$ \\
HD 6613   & $1095.0 \pm 3.3$  & $8280.0 \pm 3.5$  & $0.18 \pm 0.007$    & $8.278  \pm 0.048$ & $1.122 \pm 0.036$ & $12.521 \pm 0.039$ \\
HD 12390  & $557.85 \pm 0.34$ & $8366.68 \pm0.69$ & $0.3497 \pm 0.0038$ & $10.101 \pm 0.043$ & $2.783 \pm 0.012$ & $-11.785 \pm 0.028$ \\
HD 39923  & $1101.5 \pm 3.3$  & $8457.3 \pm 3.4$  & $0.1497 \pm 0.0086$ & $11.45  \pm 0.09$  & $-$               & $15.4 \pm 0.2$ \\
HD 55201  & $776.61 \pm 0.96$ & $8838.0 \pm 2.6$  & $0.0068 \pm 0.0065$ & $9.25  \pm 0.098$  & $-$               & $49.694 \pm 0.062$ \\
HD 147430 & $2242   \pm 32$   & $7117  \pm 33$    & $0.331  \pm 0.009 $ & $11.5  \pm 0.06$   & $0.515 \pm 0.014$ & $-31.01 \pm 0.11$  \\
HD 195543 & $674.77 \pm 0.38$ & $8758 \pm 1$      & $0.1157 \pm 0.0034$ & $10.642 \pm 0.041$ & $4.198 \pm 0.032$ & $-18.901 \pm 0.028$ \\
HD 202699 & $2077  \pm 34$    & $8721.1 \pm 2.9$  & $0.4314 \pm 0.0084$ & $11.059 \pm 0.059$ & $3.507 \pm 0.011$ & $-21.3 \pm 0.1$  \\
HD 221643 & $724.78 \pm 0.88$ & $8513.1 \pm 1.6$  & $0.1906 \pm 0.0063$ & $10.829 \pm 0.059$ & $0.441 \pm 0.035$ & $-9.305 \pm 0.052$ \\
\bottomrule
\end{tabular*}
\label{tab:orbits}
\end{table*}

We present the results of the determination of the orbital parameters with RadVel in Table~\ref{tab:orbits}, 
where we also present the uncertainties obtained with the MCMC run. 
There is one star (HD~55201) where we were not able to determine the argument of the periapsis $\omega$
because the orbit is close to circular with a low eccentricity of $e= 0.0068 \pm 0.0065$. 
The other star where we could not determine the argument of the periapsis $\omega$ was HD~39923 probably because
we do not have enough data for the long orbital period.

To maintain the readability of the paper, the detailed graphs of the RV observations and 
the fit of the RadVel toolkit 
are presented in the \ref{sec:radvelplots}.
In subplots a) we present all RV measurements during the complete observation campaign marked with yellow dots. 
The best RadVel fit is shown as a solid (blue) line. The residuals are shown in subplots b).

In Fig.~\ref{fig:HD1585}, we plotted the observed RV curve and the RadVel fit of HD~1585. This star has the lowest
semi-amplitude of the nine binary systems presented here. The recently published Gaia DR3 Part 3. Non-single stars catalog
\citep{gaianonsingle} has an entry for the star HD~1585 identifying it as an astrometric binary star. 
The orbital parameters are very similar to the ones obtained in this study. 
However, the orbital period is with $P=926.2\pm16.7$~days smaller than ours ($P=1085.8 \pm 4.4$~days).
There is a factor two of difference in the eccentricity of Gaia DR3 determined as $e=0.185\pm0.039$, 
while our eccentricity is $e=0.095 \pm 0.0088$.

The next star is HD~6613, and the RV curve is presented in Fig.~\ref{fig:HD6613}.
This star was marked as a binary by \citet{frankowski07} in a search for proper-motion binaries
in the 	Hipparcos catalog. There is also an entry in the Gaia DR3 non-single stars catalog.
They give a solution for an SB1 star, where the orbital period is $P=1090.00\pm24$~days. This coincides
within the errors with our determination of an orbital period of $P=1095.0 \pm 3.3$~days. 
The eccentricity of the Gaia DR3 solution
is $e=0.190\pm0.018$, which again agrees with our eccentricity for HD~6613 of $e=0.18 \pm 0.007$. 
In general, our determination of the orbital parameters is better because our spectrograph has 
a higher spectral resolution 
and larger wavelength range than the Gaia RV spectrometer \citep{gaiarvspectrometer}. 
The argument of periastron of the Gaia DR3 orbit is $\omega=70\pm7^\circ$ 
compared to the value of our solution of $\omega=64.3\pm2.1^\circ$.

In Fig.~\ref{fig:HD12390}, we present the complete RV measurements of the star HD~12390. 
This star is in the Catalog of ultrawide binary stars from Gaia DR2 \citep{tian20}.
However, the two stars in that Catalog have a projected physical separation of about 172,640~AU or 0.837~pc.
This is of course a much larger separation than the detected spectroscopic binary orbit of our study.
The orbital period of the spectroscopic binary is $P=557.85 \pm 0.34$~days. 
The eccentricity was found to be $e=0.3497 \pm 0.0038$ and the orbit has a semi-amplitude of $K=10.101\pm0.043$~km~s$^{-1}$.
There is no entry in the Gaia DR3 non-single stars catalog.

Figure~\ref{fig:HD39923} demonstrates the observed and fitted RV curves of the star HD~39923.
This star is also mentioned in the Gaia DR3 non-single stars catalog. 
It has an orbital solution for combined data of astrometric observations and
single-lined spectroscopy.
The orbital period given by Gaia DR3 is $P=1098.7\pm4.3$~days, while we obtained an orbital period of $P=1101.5 \pm 3.3$~days.
These values are within the uncertainties. 
For the eccentricity, Gaia DR3 found a value of $e=0.135\pm0.008$, 
which is also similar to the value of our orbital solution ($e=0.1497 \pm 0.0086$).
Despite the non-zero eccentricity, we were not able to determine the
the argument of the periapsis of the star's orbit $\omega$.

The star with the lowest eccentricity is HD~55201, and the RV curve is shown in Fig.~\ref{fig:HD55201}.
The orbit is practically circular ($e=0.0068 \pm 0.0065$). This star can also be found
in the Gaia DR3 non-single stars catalog, and it has a solution as an SB1 star. 
The eccentricity of Gaia DR3 is given with $e=0.012\pm0.025$, which also means that the orbit is circular.
The orbital period of Gaia DR3 is $P=780.1\pm4.5$~days, and our orbital period is $P=776.61 \pm 0.96$~days. 
Again, our orbital parameters are more precise and within the Gaia DR3 uncertainties.
Gaia DR3 determined a semi-amplitude of $K=8.98\pm0.13$~km~s$^{-1}$. Our semi-amplitude of the orbit
is $K=9.25\pm 0.098$~km~s$^{-1}$.
It is also worth mentioning that this star has also an entry in the Gaia DR3 non-single stars 
catalog for sources having a non-linear proper motion.

Figure~\ref{fig:HD147430} demonstrates the observed and fitted RV curve of the star HD~147430.
This star has the longest period of $P=2242\pm 32$~days,
and it will be necessary to obtain more observations to determine the 
orbital parameters more precisely, because so far we have not covered a full orbital period. 
This star is mentioned in the Gaia DR3 non-single stars catalog
as having a non-linear proper motion, but there is no orbital solution to which we could
compare our results. The eccentricity is $e=0.331\pm 0.009$, and
the semi-amplitude is $K=11.5\pm 0.06$~km~s$^{-1}$.

The full RV curve of HD~195543 is shown in Fig.~\ref{fig:HD195543}. 
This star is marked as a spectroscopic binary in SIMBAD.
It is in the Gaia DR3 non-single stars catalog and has a combined astrometric + 
single-lined spectroscopic orbital solution.
We determined an orbital period of $P=674.77 \pm 0.38$~days, which is similar within
the uncertainties to the value of the Gaia DR3 orbital solution ($P=674.27\pm0.96$~days). 
The eccentricity given by Gaia DR3 is $e=0.101\pm0.005$, while we determined 
a value of $e=0.1157 \pm 0.0034$ for our orbital solution.

The observed and fitted RV curves of HD~202699 are presented in Fig.~\ref{fig:HD202699}. 
The orbital period of $P=2077 \pm 34$~days is quite long, 
and we have not covered a full orbital period with our observations.
This star has the most eccentric orbit of the nine binaries
with a value for the eccentricity of $e=0.4314\pm0.0084$.
This star does not have an entry in the Gaia DR3 non-single stars catalog.

Figure~\ref{fig:HD221643} demonstrates the observed and fitted RV curves 
of the star HD~221643.
This star shows an interesting bump in the observed RV curve around JD~2458750.
It has an almost exact two-year orbital period, this means that we have observed the star
always during the same phases of the orbit.
This star has an entry in the Gaia DR3 non-single stars catalog 
with an orbital solution as an astrometric binary. 
The orbital period is given as $P=725.19\pm8.6$~days, while we determined
a value of $P=724.78 \pm 0.88$~days,
which is within the Gaia DR3 uncertainty. The eccentricity given by
Gaia DR3 has a value of $e=0.195\pm0.062$, and the value for
the eccentricity of our orbital solution is $e=0.1906 \pm 0.0063$.

%
%
% Spectral analysis, evolution tracks
%
\section{Stellar parameters}\label{sec:param}

Using the intermediate resolution spectra ($R\approx 20,000$),
we determined all stellar parameters of the primary stars and
placed them in the Hertzsprung-Russell diagram (HRD), which
allowed us to obtain the masses and ages.

\subsection{Distances, extinction, and absolute magnitudes in the $V$-band}

\begin{table*}[t]
\centering
\caption{Observational properties of nine spectroscopic binary stars. 
The extinction ($E(B-V)$) was determined with the 3D dust mapping to obtain the absolute magnitudes
in the $V$-band $M_V$.}
\tabcolsep=0pt
\begin{tabular*}{500pt}{@{\extracolsep\fill}lcccccccc@{\extracolsep\fill}}
\toprule
\textbf{Star} & \textbf{Spectral type}  & $m_V$ & \textbf{parallax (mas)} & \textbf{distance (pc)} & $E(B-V)$ & $M_V$ \\
\midrule
HD 1585   & K0 & $6.782\pm0.001$ & $2.87\pm0.21$  & $352.3\pm22.7$ & $0.020\pm0.02$ & $-1.01\pm0.16$ \\  
HD 6613   & G5 & $6.99\pm0.01$   & $3.35\pm0.14$  & $301.9\pm9.3$  & $0.030\pm0.03$ & $-0.50\pm0.12$ \\  
HD 12390  & K0 & $6.54\pm0.01$   & $2.66\pm0.10$  & $380.9\pm9.3$  & $0.060\pm0.03$ & $-1.55\pm0.11$ \\   
HD 39923  & K5 & $7.63\pm0.01$   & $2.77\pm0.33$  & $366.2\pm44.5$ & $0.145\pm0.04$ & $-0.64\pm0.31$ \\
HD 55201  & G5 & $7.29\pm0.01$   & $5.58\pm0.17$  & $180.4\pm4.5$  & $0.000\pm0.02$ & $ 1.01\pm0.08$ \\ 
HD 147430 & K2 & $6.920\pm0.009$ & $3.59\pm0.12$  & $280.6\pm7.2$  & $0.115\pm0.03$ & $-0.68\pm0.11$ \\ 
HD 195543 & K5 & $7.36\pm0.01$   & $2.09\pm0.06$  & $484.6\pm8.9$  & $0.030\pm0.03$ & $-1.16\pm0.10$ \\ 
HD 202699 & K2 & $7.00\pm0.01$   & $3.02\pm0.10$  & $335.0\pm8.1$  & $0.078\pm0.03$ & $-0.87\pm0.11$ \\ 
HD 221643 & K5 & $7.11\pm0.01$   & $2.28\pm0.13$  & $444.1\pm20.1$ & $0.065\pm0.04$ & $-1.33\pm0.16$ \\ 
\bottomrule
\end{tabular*}
\label{starlist}
\end{table*}

In Table~\ref{starlist}, we present the observational properties like spectral type and
apparent magnitude in the $V$-band ($m_V$) of the stars. The information was obtained from
the SIMBAD database, and the parallaxes were taken from the Gaia~DR3 archive.
We already discussed in Paper~I that the binarity may cause a systematic effect on the parallaxes.
The presented distances include the zero-point correction for Gaia DR3 parallaxes \citep{gaiaedr32}.
Because some stars are located at considerable distances,
we estimated the reddening of the stars using the 3D dust mapping \citep{green19}(\url{http://argonaut.skymaps.info/}) and
obtained the color excess $E(B-V)$ using the dust map at the given positions and Gaia DR3 distances.
Assuming an extinction law of $R(V)=3.1$, the corrections for the $V$-band ($A_V$) magnitude were 
determined. We then determined the absolute magnitudes in the $V$-band ($M_V$).
No star has a specification of the luminosity class in SIMBAD, but it is clear from the absolute magnitudes
that all stars must be in the giant phase.
Before we could continue with the determination of the luminosities,
it was necessary to apply bolometric corrections (BC),
for which we needed exact stellar parameters like the effective temperature $T_\mathrm{eff}$. 
Therefore, we needed to perform a detailed analysis of the observed spectra.

%
% Spectral analysis
%
\subsection{Spectral analysis with iSpec}

As discussed in the previous papers, we obtained spectra with a low signal-to-noise ratio (S/N) 
of about 30 in the red channel of the HEROS spectrograph. 
Therefore, we combined ten spectra with the highest S/N to
obtain one spectrum with an S/N that allows a determination of the stellar parameters
from spectral analysis.

For the determination of the stellar parameters, we used
the same method as described in Paper~II. 
That is, the spectral analysis toolkit iSpec \citep{Blanco2014} in its
Python 3 version (v2020.10.01) \citep{Blanco2019}.
We applied the improved method established in \citet{rosas21} and used in Paper~II,
which works very well with giant stars. Please consult Paper~II for
a detailed description of this method.

\begin{table*}[t]%
\centering
\caption{Basic stellar parameters of the nine spectroscopic binaries determined with spectral analysis using iSpec.}
\tabcolsep=0pt%
\begin{tabular*}{500pt}{@{\extracolsep\fill}lccccccc@{\extracolsep\fill}}
\toprule
\textbf{Star} & $T_\mathrm{eff}$~(K)  & $\log{g}$ & $[M/H]$ & $[\alpha/Fe]$ & $v_\mathrm{mic}$~(km~s$^{-1}$) &
$v_\mathrm{mac}$~(km~s$^{-1}$) & $v_\mathrm{rot}\sin{i}$~(km~s$^{-1}$) \\
\midrule
HD 1585   & $3962\pm 100$ & $1.04\pm0.05$ & $-0.36\pm0.02$  & $-0.11\pm0.03$ & $1.74\pm0.03$ & $2.99\pm0.62$ & $3.16\pm1.01$\\ 
HD 6613   & $5110\pm 100$ & $2.82\pm0.11$ & $0.03\pm0.04$   & $-0.06\pm0.05$ & $1.65\pm0.05$ & $5.13\pm0.70$ & $3.42\pm1.05$\\
HD 12390  & $4803\pm 100$ & $1.74\pm0.10$ & $-0.13\pm0.03$  & $-0.03\pm0.04$ & $1.75\pm0.05$ & $3.79\pm0.82$ & $5.30\pm0.00$\\
HD 39923  & $3964\pm 100$ & $1.18\pm0.12$ & $-0.30\pm0.04$  & $0.03\pm0.05$ &  $1.52\pm0.05$ & $3.92\pm0.84$ & $1.89\pm0.00$\\
HD 55201  & $4721\pm 100$ & $2.62\pm0.05$ & $0.12\pm0.02$   & $-0.08\pm0.03$ & $1.59\pm0.03$ & $4.21\pm0.74$ & $0.00\pm0.00$\\
HD 147430 & $4310\pm 100$ & $1.53\pm0.04$ & $-0.31\pm0.02$  & $0.02\pm0.03$  & $1.22\pm0.03$ & $2.38\pm0.54$ & $4.91\pm1.63$\\
HD 195543 & $4199\pm 100$ & $1.31\pm0.04$ & $-0.32\pm0.02$  & $-0.01\pm0.03$ & $1.74\pm0.03$ & $3.76\pm0.54$ & $2.06\pm1.63$\\
HD 202699 & $4170\pm 100$ & $1.11\pm0.04$ & $-0.32\pm0.02$  & $0.06\pm0.03$  & $1.60\pm0.03$ & $3.71\pm0.54$ & $3.83\pm1.63$\\
HD 221643 & $3945\pm 100$ & $1.46\pm0.04$ & $-0.11\pm0.02$  & $-0.09\pm0.03$ & $1.68\pm0.03$ & $3.33\pm0.54$ & $3.89\pm1.63$\\
\bottomrule
\end{tabular*}
\label{tab:param}
\end{table*}

We present the results of the stellar parameter determination in Table~\ref{tab:param}.
The effective temperatures range from about 3900 to 5100~K,
and the values for the surface gravities indicate that all stars are in the giant phase. 
The rotational velocities are difficult to determine and include a factor of $\sin{i}$ 
because of the inclination of the stellar rotation axis.
Below, we will give an estimate of the orbital inclinations $i$ of the binary system orbits,
which are usually similar to the inclinations of the rotation axis of the primary stars.
However, the intermediate spectral resolution of the HEROS spectrograph does not really allow a physically meaningful determination
of the turbulence and rotational velocities. For all parameters that are determined with line profiles, it is necessary to use
high-quality (high S/N) spectra for a reliable determination.

Table~\ref{tab:param} gives also the fitting uncertainties of iSpec.
These are very low and do not represent the uncertainties that are introduced from observations, models 
or spectral line data for the fitting process. To continue with our parameter determination,
we assumed a conservative value of 100~K for the uncertainty of the effective temperatures.

\subsection{Masses and ages}

\begin{table*}[t]%
\centering
\caption{The masses and ages of the five primary stars of the spectroscopic binaries determined by comparing stellar evolution tracks to the positions in the HRD.}
\tabcolsep=0pt%
\begin{tabular*}{500pt}{@{\extracolsep\fill}lccccccc@{\extracolsep\fill}}
\toprule
\textbf{Star} & $M_V$ & BC & $M_\mathrm{bol}$ & $\log{T_\mathrm{eff}} $ & $\log{L/L_\odot} $ & \textbf{mass~$(M_\odot)$}  & \textbf{age~(Myr)} \\% masses and ages
\midrule
\rule{0pt}{10pt}HD 1585  & $-1.01\pm0.16$  & $-1.078 \pm 0.014$ & $-2.03 \pm 0.16$ & $3.598\pm 0.011$ & $2.73\pm0.07$ & $1.40\pm 0.20$ & $3184^{+2249}_{-1100}$ \\
\rule{0pt}{12pt}HD 6613  & $-0.50\pm0.12$  & $-0.299 \pm 0.011$ & $-0.71 \pm 0.12$ & $3.708\pm 0.009$ & $2.20\pm0.05$ & $3.35^{+0.05}_{-0.10}$ & $291^{+25}_{-12}$ \\
\rule{0pt}{12pt}HD 12390 & $-1.55\pm0.11$  & $-0.509 \pm 0.012$ & $-1.87 \pm 0.11$ & $3.682\pm 0.009$ & $2.67\pm0.05$ & $4.00^{+0.40}_{-0.20}$ & $195^{+48}_{-51}$ \\
\rule{0pt}{12pt}HD 39923 & $-0.64\pm0.31$  & $-1.458 \pm 0.016$ & $-1.65 \pm 0.31$ & $3.598\pm 0.011$ & $2.58\pm0.17$ & $1.15\pm 0.25$ & $6309^{+9108}_{-3132}$ \\
\rule{0pt}{12pt}HD 55201 & $ 1.01\pm0.08$  & $-0.363 \pm 0.011$ & $0.65 \pm 0.08$  & $3.674\pm 0.009$ & $1.66\pm0.04$ & $1.65\pm 0.20$ & $2152^{+996}_{-624}$ \\
\rule{0pt}{12pt}HD 147430 & $-0.68\pm0.11$ & $-0.993 \pm 0.014$ & $-1.31 \pm 0.11$ & $3.634\pm 0.010$ & $2.45\pm0.05$ & $1.60\pm 0.25$ & $2075^{+1522}_{-720}$ \\
\rule{0pt}{12pt}HD 195543 & $-1.16\pm0.10$ & $-0.831 \pm 0.014$ & $-1.90 \pm 0.10$ & $3.623\pm 0.010$ & $2.68\pm0.05$ & $1.80\pm 0.20$ & $1469^{+611}_{-114}$ \\
\rule{0pt}{12pt}HD 202699 & $-0.87\pm0.11$ & $-1.009 \pm 0.014$ & $-1.63 \pm 0.11$ & $3.620\pm 0.011$ & $2.57\pm0.05$ & $1.50\pm 0.25$ & $2541^{+2169}_{-945}$ \\
\rule{0pt}{12pt}HD 221643 & $-1.33\pm0.16$ & $-1.244 \pm 0.016$ & $-2.37 \pm 0.16$ & $3.596\pm 0.011$ & $2.87\pm0.09$ & $2.15\pm 0.35$ & $1262^{+440}_{-440}$ \\
\bottomrule
\end{tabular*}
\label{tab:massage}
\end{table*}
We determined the BCs with the published code presented in \citet{casagrande14,casagrande18}.
This allowed us to obtain the absolute bolometric magnitudes $M_\mathrm{bol}$ and
luminosities $L$ (in solar luminosities $L_\odot$) of the stars. 
We present the results in Table~\ref{tab:massage}.
Using the luminosities and effective temperatures of the primary stars,
we could finally place them in the HRD.
We computed stellar evolution tracks with the Cambridge (UK) Eggleton code
in its updated version \citep{pols97,pols98} and determined
the masses and ages of the primary stars.
The results are also presented in Table~\ref{tab:massage}.
The masses ranged from 1.15 to 4.0 $M_\odot$, and the ages from 195~Myr to $\approx 6.3$~Gyr.
In general, masses and especially ages of stars are difficult to determine 
because the evolution tracks are very close together during the giant phase.
\begin{figure}[t]
    \centerline{\includegraphics[width=0.5\textwidth]{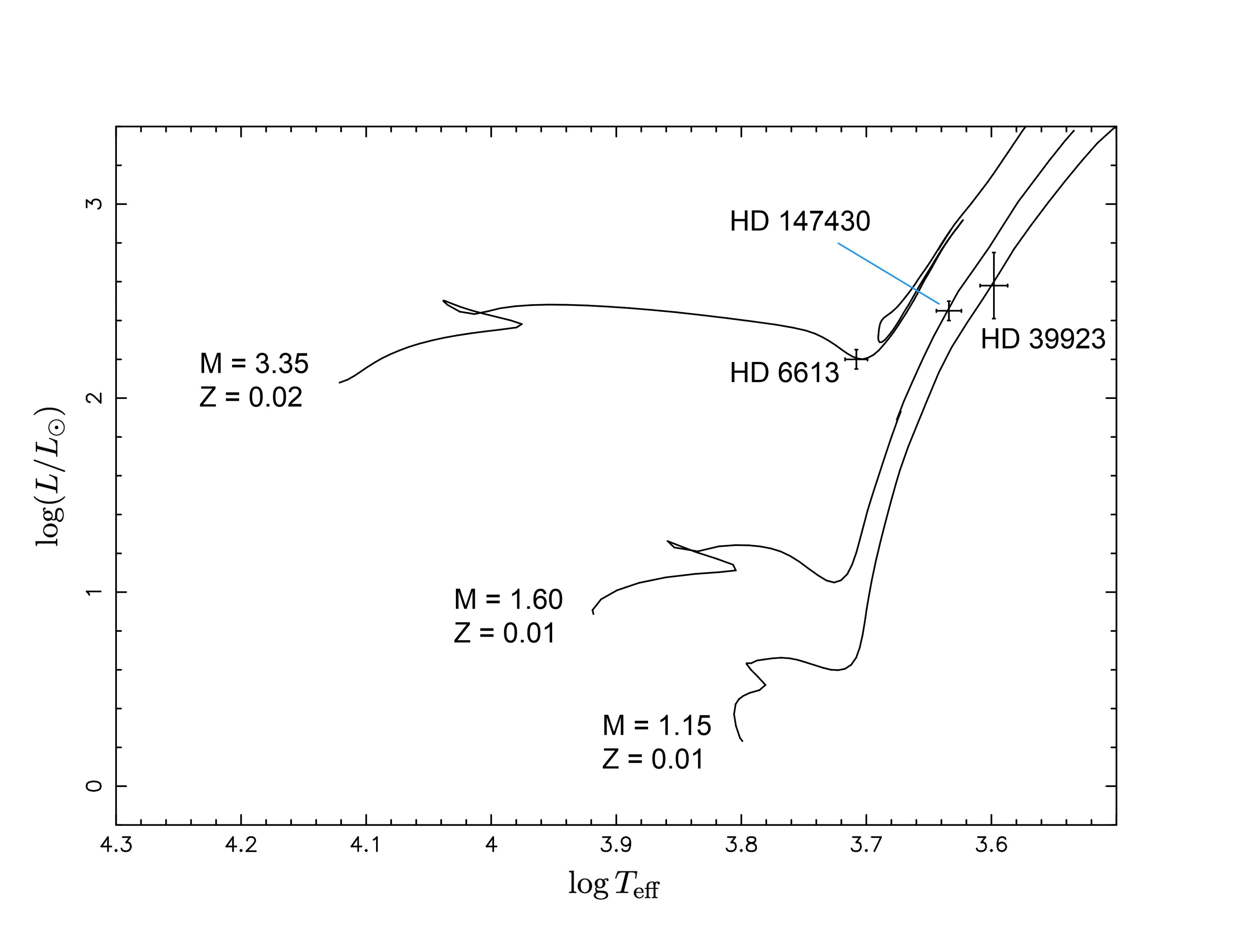}}
        \caption{Positions of the primary components of HD 6613, HD 39923, and HD 147430 in the HRD and the best fitting stellar evolution tracks (solid lines) of the respective masses using
        tracks with the mentioned value of the metallicity $z$.
\label{fig:tracks}}
\end{figure}
\begin{figure}[t]
    \centerline{\includegraphics[width=0.5\textwidth]{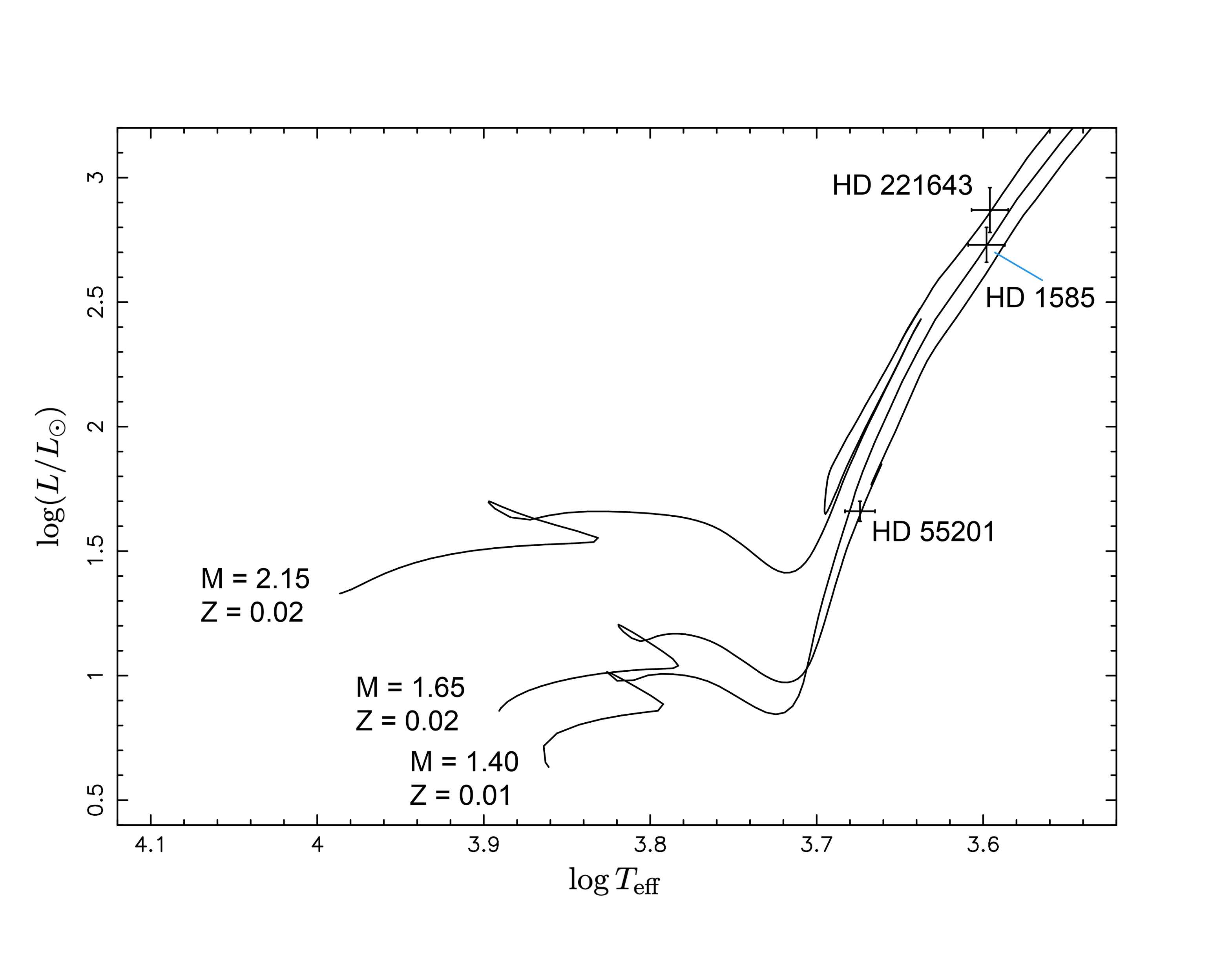}}
        \caption{Positions of the primary components of HD 1585, HD 55201, and HD 221643 in the HRD and the best fitting stellar evolution tracks (solid lines) of the respective masses
        using tracks with the mentioned value of the metallicity $z$.
\label{fig:tracks2}}
\end{figure}
\begin{figure}[t]
    \centerline{\includegraphics[width=0.5\textwidth]{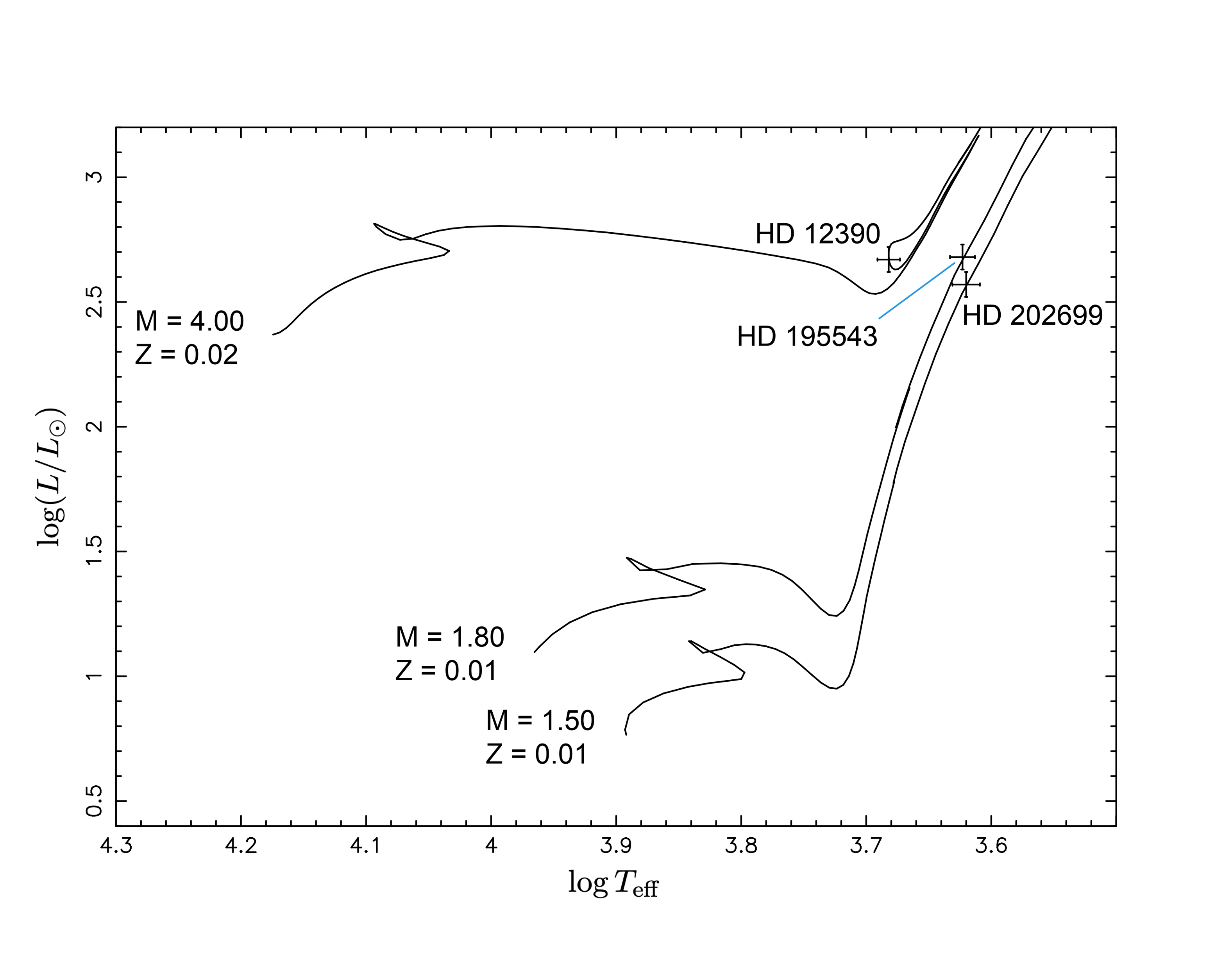}}
        \caption{Positions of the primary components of HD 12390, HD 195543, and HD 202699 in the HRD and the best fitting stellar evolution tracks (solid lines) of the respective masses
        using tracks with the mentioned value of the metallicity $z$.
\label{fig:tracks3}}
\end{figure}
In Fig.~\ref{fig:tracks}, 
we show the positions of the stars HD~6613, HD~39923, and HD~147430 in the theoretical HRD with stellar evolution tracks.
The HRD positions of three other stars (HD~1585, HD~55201, and HD~221643) are presented in Fig.~\ref{fig:tracks2}.
The remaining three stars HD~12390, HD~195543, and HD~202699 and their positions in the HRD are shown in Fig.~\ref{fig:tracks3}.
We used two different values for the metallicity $z$ for the tracks. The used value
is indicated for each track.

We performed a consistency check, where we calculated the parallax-related surface gravities of all stars as described in \citet{rosas21}
and \citet{paper2}.
Knowing the luminosity $L$, the mass $M_1$, and the distance from Gaia DR3, we calculated the surface gravity using
basic relations between the stellar parameters. We found that the two
independently determined surface gravities are very similar. The largest difference
was found for HD~202699 with $\Delta \log{g}\approx0.35$.

\subsection{Secondary star masses and orbital inclinations $i$}

\begin{table*}[t]%
\centering
\caption{The mass function ($f$) of the nine spectroscopic binary systems, which we used to
determine the minimum mass of the secondary stars ($m_2$). Some inclinations $i$ were determined. Other values
were taken from the Gaia DR3 data.}
\tabcolsep=0pt%
\begin{tabular*}{500pt}{@{\extracolsep\fill}lcccccccccc@{\extracolsep\fill}}
\toprule
\textbf{Star} & $f$ ($M_\odot$) & $M_1$ ($M_\odot$) & $m_2=M_2\sin{i}$ ($M_\odot$) & $m_2^\dagger\sqrt{r}$ ($M_\mathrm{Jup}$) & $\gamma$ & $i$  & $M_2$ ($M_\odot$) \\
\midrule
\rule{0pt}{10pt}HD 1585 &  $0.0467\pm0.0011$  & $1.40\pm 0.20$         & $0.47\pm0.04$  & $283^{+81}_{-36}$ & $0.484$ & $50.7^\circ ~^{+7.0^\circ}_{-5.6^\circ}$  & $0.60\pm0.1$ \\  
\rule{0pt}{12pt}HD 6613 &  $0.0613\pm0.0010$  & $3.35^{+0.05}_{-0.10}$ & $0.90\pm0.03$  & $474^{+133}_{-53}$& $0.491$ & $52.9^\circ ~^{+6.1^\circ}_{-5.6^\circ}$  & $1.13\pm0.1$ \\
\rule{0pt}{12pt}HD 12390 & $0.0490\pm0.0005$  & $4.00^{+0.40}_{-0.20}$ & $0.94\pm0.05$  & $176^{+62}_{-43}$ & $0.155$ & $55.7^\circ ~^{+10.0^\circ}_{-6.9^\circ}$ & $1.14\pm0.1$ \\
\rule{0pt}{12pt}HD 39923 & $0.16946\pm0.0011$ & $1.15\pm 0.25$         & $0.67\pm0.02$                       &    &   & $44.92^\circ\pm1.17^\circ$                & $0.95\pm0.03$ \\ %inclination is from Gaia DR3 solution
\rule{0pt}{12pt}HD 55201 & $0.0637\pm0.0021$  & $1.65\pm 0.20$         & $0.58\pm0.05$  & $327^{+91}_{-35}$ & $0.157$ & $30.6^\circ ~^{+8.7^\circ}_{-3.9^\circ}$  & $1.14\pm0.02$ \\
\rule{0pt}{12pt}HD 147430 & $0.2969\pm0.0059$ & $1.60\pm 0.25$         & $1.02\pm0.10$  & $679^{+190}_{-73}$& $0.860$ & $50.1^\circ ~^{+6.3^\circ}_{-5.4^\circ}$  & $1.33\pm0.1$ \\
\rule{0pt}{12pt}HD 195543 & $0.0826\pm0.0010$ & $1.80\pm 0.20$         & $0.67\pm0.03$                       &    &   & $74.0^\circ\pm6.9^\circ$                  & $0.70\pm0.03$ \\ %inclination is from Gaia DR3 solution
\rule{0pt}{12pt}HD 202699 & $0.2154\pm0.0048$ & $1.50\pm 0.25$         & $0.86\pm0.09$  & $649^{+182}_{-70}$& $0.841$ & $48.4^\circ ~^{+6.6^\circ}_{-5.4^\circ}$  & $1.16\pm0.10$ \\
\rule{0pt}{12pt}HD 221643 & $0.0902\pm0.0010$ & $2.15\pm 0.35$  	   & $0.78\pm0.03$                       &    &   & $57.7^\circ\pm 2.6^\circ$                 & $0.92\pm0.03$ \\ %inclination is from Gaia DR3 solution
\bottomrule
\end{tabular*}
\label{tab:masses}
\end{table*}

We have now the masses of the primary components from the fits of
evolution tracks to the HRD positions. Thus, we now attempt to estimate the
masses of the secondary components and the inclinations.

The Kepler's laws provide a formula that connects the masses of the
two components of a binary system with three observable orbital parameters,
which are the orbital period $P$, the semi-amplitude $K$, and the eccentricity $e$.
This function $f$ is called the mass function and can be calculated by the formula
\begin{equation}
\label{eq1}
f=\frac{M_2^3\sin^3{i}}{(M_1+M_2)^2}=\frac{P K^3}{2\pi G}(1-e^2)^{3/2},
\end{equation}
where $G$ represents the gravitational constant. 
This formula can be used to calculate the masses of both stars from one spectroscopic orbit.
However, this is only possible
if both inclination $i$ and the mass ratio are known, which is only the case for eclipsing SB2
systems (i.e., $\sin{i}$ close to 1).
In some cases, the inclination $\sin{i}$ can be obtained from an astrometric orbit. 
We were only able to calculate the minimal masses of the secondary stars ($m_2=M_2\sin{i}$) 
solving Eq.~\ref{eq1}.
We present the results in Table~\ref{tab:masses}.

Some of the binary stars were also identified as binaries in an 
analysis of the proper motion anomaly of Hipparcos stars using Gaia DR2 data \citep{kervella19}.
They present masses of secondary stars normalized for a 1~AU circular orbit $m_2^\dagger$, 
because the orbital distance ($r$ in AU) from the primary was unknown to them.
This normalized mass is related to the minimal mass of the secondary 
by the relation $m_2^\dagger\sqrt{r}$.
There is also a sensitivity factor $\gamma$, that
increased the actual signal and, therefore, the normalized minimal secondary masses $m_2^\dagger$.
As described in Paper~II, we can combine the results of \citet{kervella19} with ours
and determine the orbital inclination $i$ using the equation
\begin{equation}
i=\arctan{\frac{m_2}{m_2^\dagger\sqrt{r}}}.
\end{equation}
We were then able to calculate the actual mass of the secondary with the equation $M_2=m_2/\sin{i}$. 

In Table~\ref{tab:masses}, we present the factor $\gamma$ and 
the results for the inclinations $i$ and the secondary masses $M_2$. 
All orbital inclinations $i$ are around $50^\circ$. The lowest inclination
has the star HD~55201 ($i=30.6^\circ ~^{+8.7^\circ}_{-3.9^\circ}$).

The value of the inclination $i$ for HD 1585, from an astrometric orbital solution of Gaia DR3 non-single stars catalog
is $48.9^\circ\pm2.4^\circ$. Our inclination is $50.7^\circ ~^{+7.0^\circ}_{-5.6^\circ}$, which agrees within the errors.
This shows that our approach to determining the inclinations with the data of \citet{kervella19} worked quite well.

There are three stars (HD~39923, HD~195543, and HD~221643) where we did not have data from \citet{kervella19},
but where values for the inclinations exist in the Gaia~DR3 non-single stars catalog. 
Thus, we used the inclinations from Gaia~DR3 to calculate the masses of the companion stars ($M_2$).

As can be seen in Table~\ref{tab:masses}, the secondary stars have lower masses and are, therefore, main sequence stars.
Comparing their luminosities to the luminosities of the giant primary stars, one can clearly see that the luminosity ratio is
at least a factor of 200. Thus, it was not possible for us to detect any signal from the
secondary components in our observed spectra.

%
% Summary
%
\section{Summary}\label{sec:con}

We determined the orbital parameters of nine well-separated bright spectroscopic binaries, 
which have orbital periods of $P>500$~days. We were able to compare some of our orbital solutions
with solutions published in the  Gaia DR3 non-single stars catalog.
They agree within the uncertainties, but our orbital solutions are more precise
because of the higher resolution of our spectrograph. 
We used the intermediate-resolution spectra of the HEROS spectrograph
and applied our automatic stellar parameter determination script based on iSpec.
With corrections for interstellar extinction and BCs, we placed
the stars in the HRD and compared the locations with stellar evolution tracks that we
calculated with the Eggleton code.
This allowed us to determine the masses and ages of the primary stars.
We calculated the minimal masses of the secondary and applied
a method of comparison with results from the analysis of proper motions anomaly of \citet{kervella19}.
This allowed us the determine the inclinations and, therefore, exact masses of most of the secondary
stars in the binary systems.

The presented thorough analysis of nine well-separated spectroscopic binaries, which contain a giant,
and the determination of the masses of both components
is of good help to quantitatively test evolution models.
It is also worth mentioning that the age of both components is the same as
the stars formed at the same time. 
This puts further constraints on any further evolution analysis of the binary systems.

\section*{Acknowledgments}

We thank the University of Guanajuato for the grant for the project 057/2023
of the {\it Convocatoria Institucional de Investigaci\'on Cient\'ifica 2023}.
This work has made use of data from the European Space Agency (ESA) mission
{\it Gaia} (\url{https://www.cosmos.esa.int/gaia}), processed by the {\it Gaia}
Data Processing and Analysis Consortium (DPAC,
\url{https://www.cosmos.esa.int/web/gaia/dpac/consortium}). Funding for the DPAC
has been provided by national institutions, in particular the institutions
participating in the {\it Gaia} Multilateral Agreement.
This work has made use of the VALD database, operated at Uppsala University, 
the Institute of Astronomy RAS in Moscow, and the University of Vienna.
This research has made use of the SIMBAD database, operated at CDS, Strasbourg, France.
This research has made use of the VizieR catalog access tool, CDS, Strasbourg, France \citep{vizier}.

% References
\bibliography{all}

%\section*{Author Biography}
%\begin{biography}{}{\textbf{Dennis Jack.} is a full professor at the University of Guanajuato. He obtained his Diploma and PhD at the
%University of Hamburg. In 2012, he moved to Mexico and works on various topics in the field of stellar astrophysics.}
%\end{biography}

%%%%%%%%%%%%%%%%% APPENDICES %%%%%%%%%%%%%%%%%%%%%
\appendix
\section{RadVel plots}
\label{sec:radvelplots}

We present in this section the plots of the RV orbits of all nine spectroscopic binaries.

\begin{figure*}[t]
    \centerline{\includegraphics[width=0.9\textwidth]{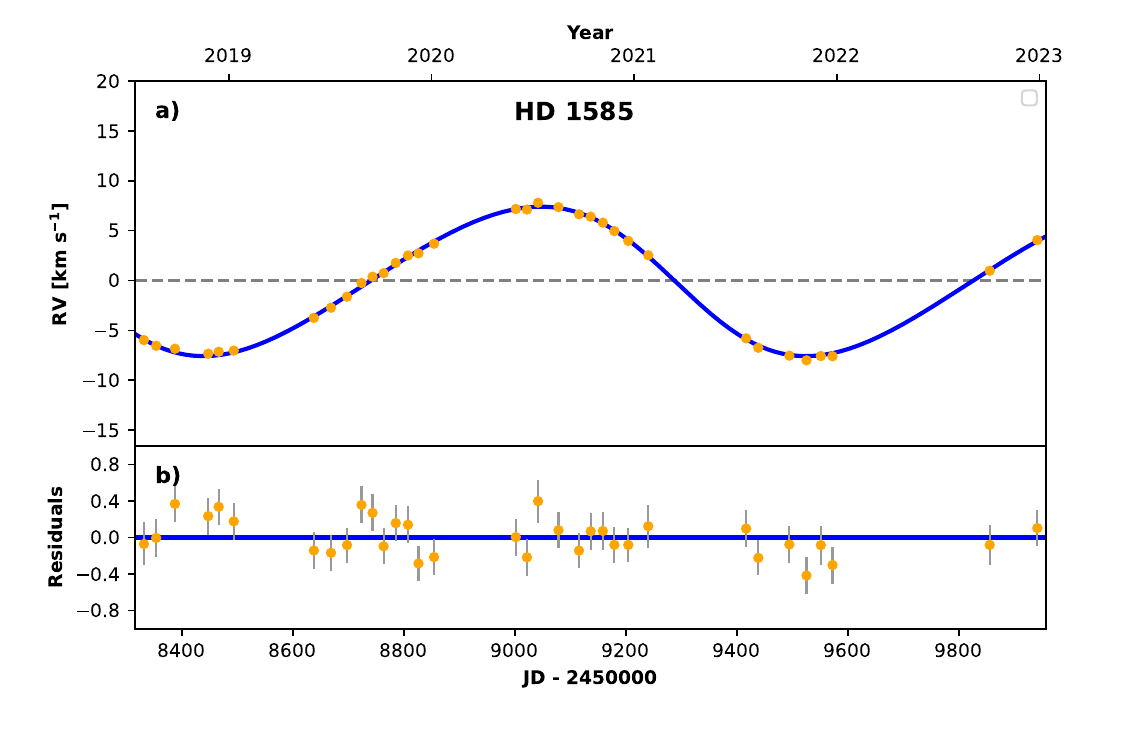}}
        \caption{RV curve of HD~1585: a) the complete measurements (circles) and RadVel fit (solid line),
        b) the residuals.\label{fig:HD1585}}
\end{figure*}

\begin{figure*}[t]
    \centerline{\includegraphics[width=0.9\textwidth]{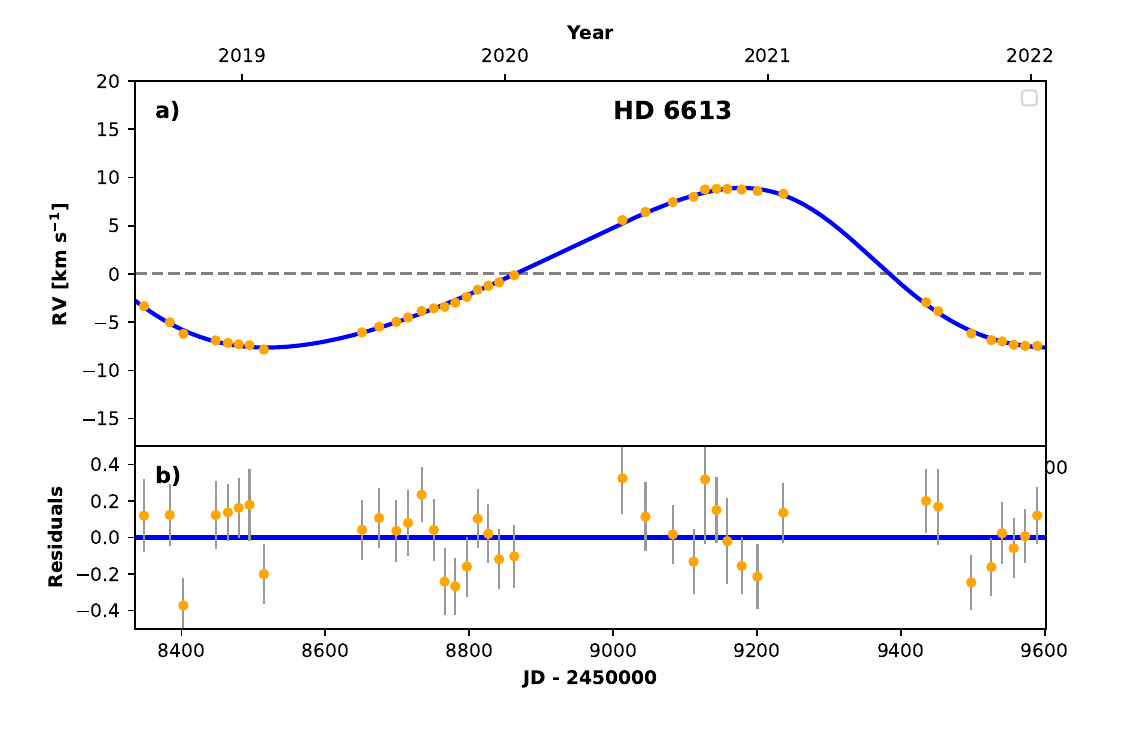}}
        \caption{RV curve of HD~6613: a) the complete measurements (circles) and RadVel fit (solid line),
        b) the residuals.\label{fig:HD6613}}
\end{figure*}

\begin{figure*}[t]
    \centerline{\includegraphics[width=0.9\textwidth]{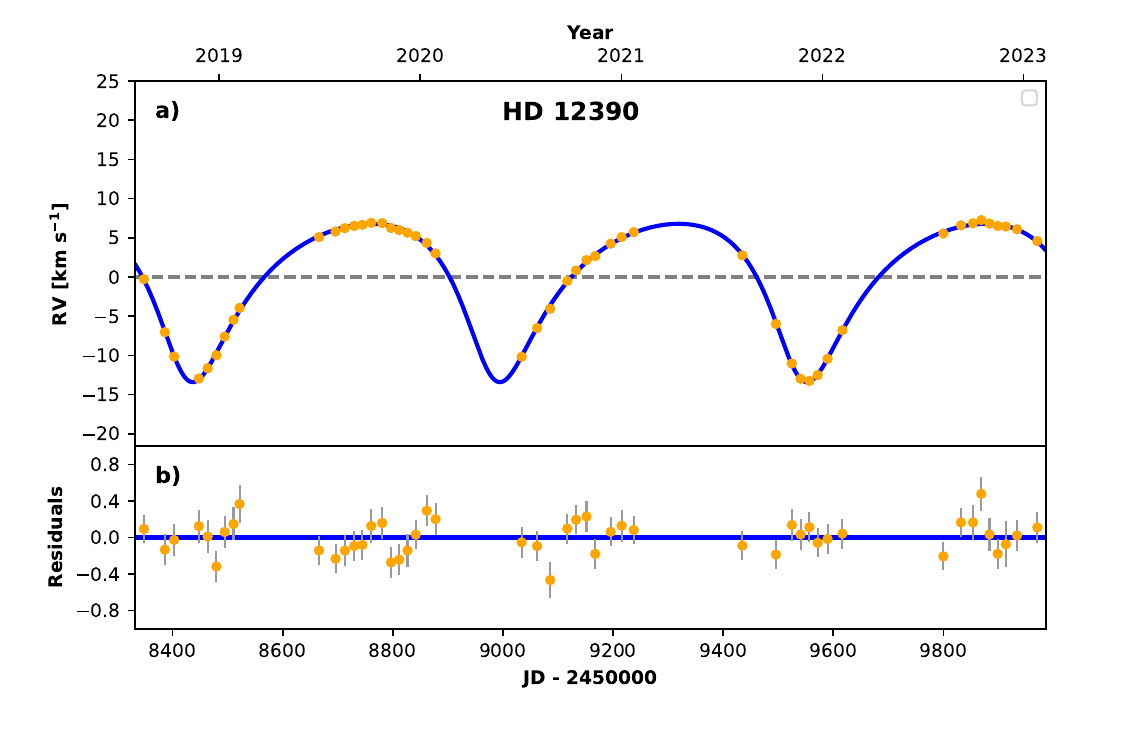}}
        \caption{RV curve of HD~12390: a) the complete measurements (circles) and RadVel fit (solid line),
        b) the residuals.\label{fig:HD12390}}
\end{figure*}

\begin{figure*}[t]
    \centerline{\includegraphics[width=0.9\textwidth]{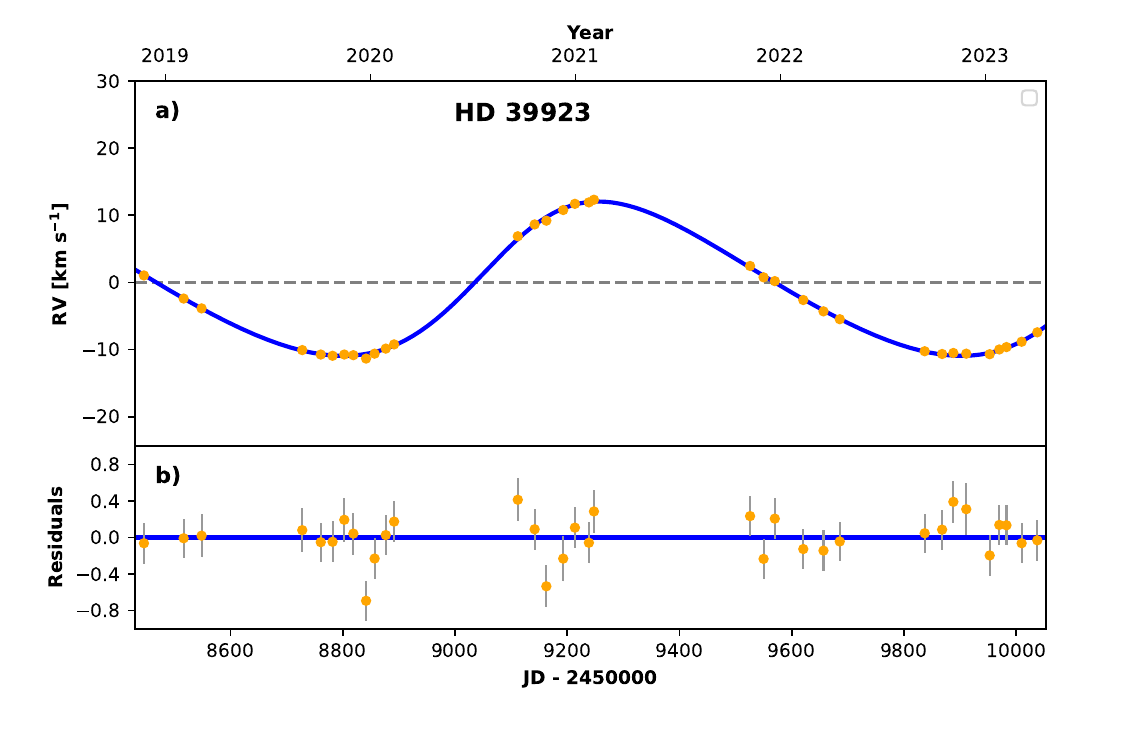}}
        \caption{RV curve of HD~39923: a) the complete measurements (circles) and RadVel fit (solid line),
        b) the residuals.\label{fig:HD39923}}
\end{figure*}
\begin{figure*}[t]
    \centerline{\includegraphics[width=0.9\textwidth]{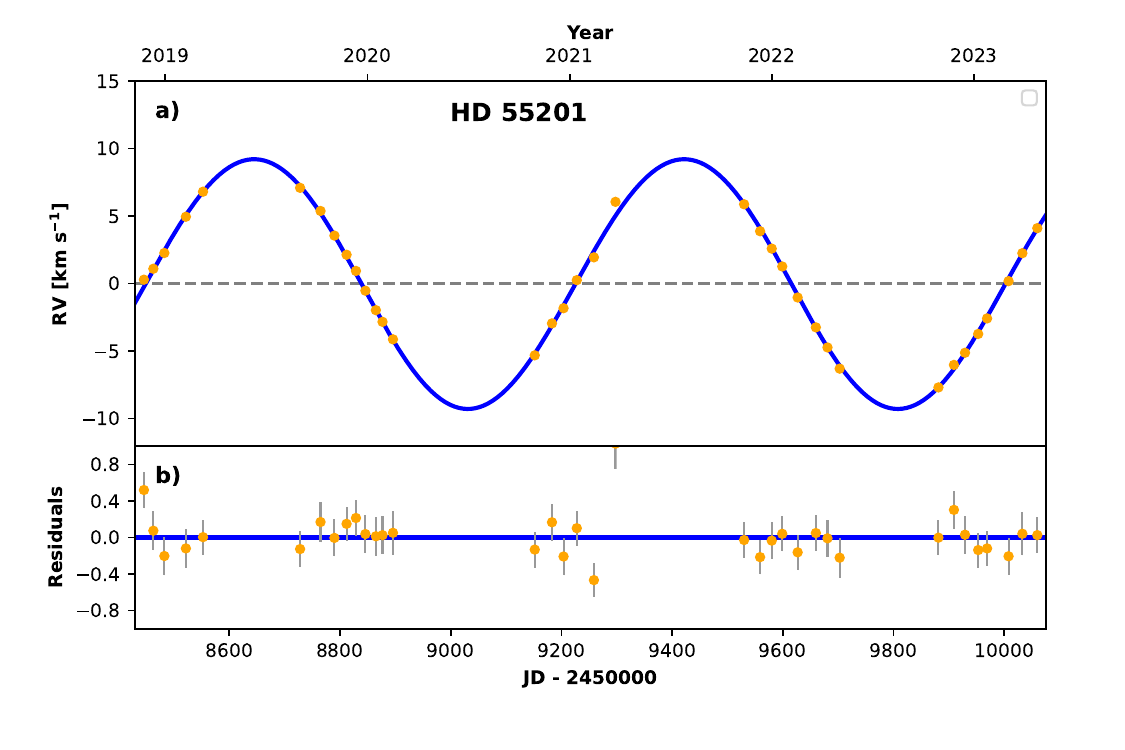}}
        \caption{RV curve of HD~55201: a) the complete measurements (circles) and RadVel fit (solid line),
        b) the residuals.\label{fig:HD55201}}
\end{figure*}

\begin{figure*}[t]
    \centerline{\includegraphics[width=0.9\textwidth]{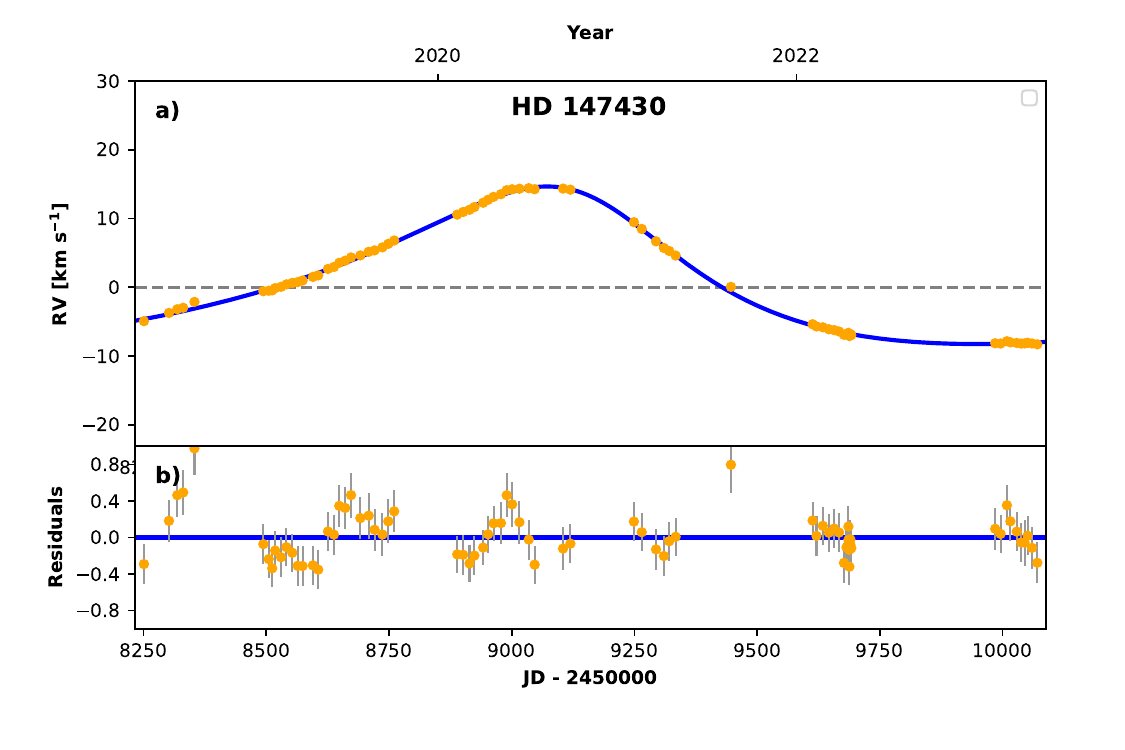}}
        \caption{RV curve of HD~147430: a) the complete measurements (circles) and RadVel fit (solid line),
        b) the residuals.\label{fig:HD147430}}
\end{figure*}

\begin{figure*}[t]
    \centerline{\includegraphics[width=0.9\textwidth]{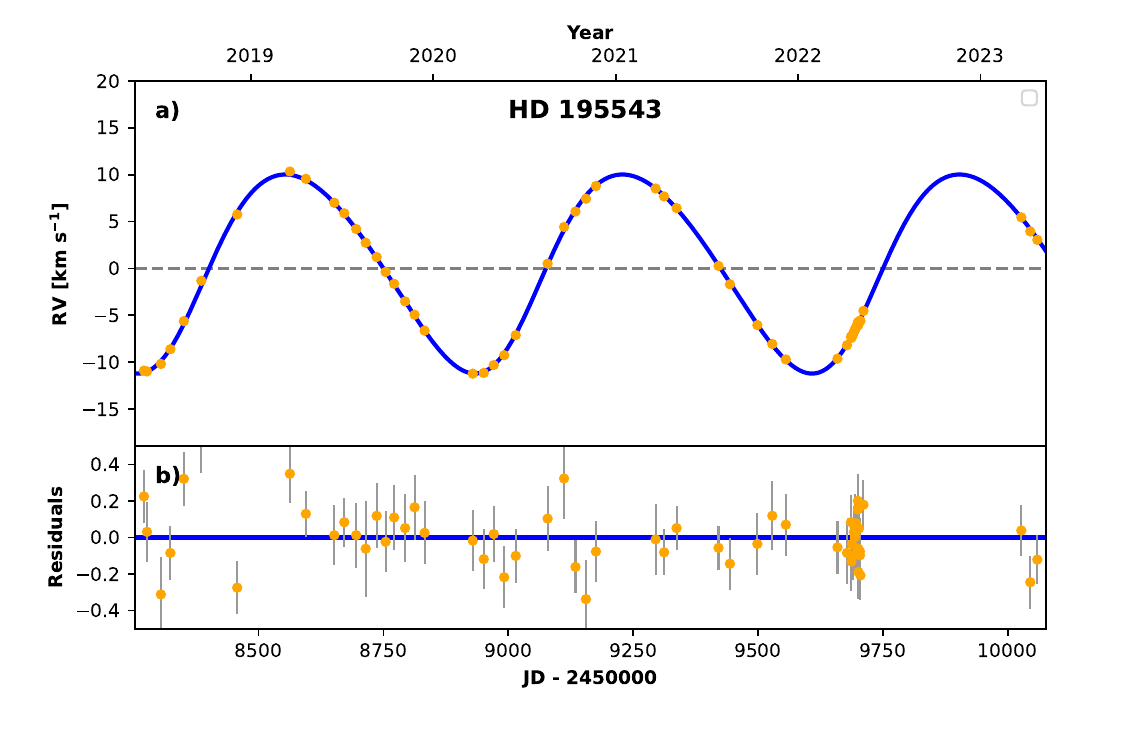}}
        \caption{RV curve of HD~195543: a) the complete measurements (circles) and RadVel fit (solid line),
        b) the residuals.\label{fig:HD195543}}
\end{figure*}

\begin{figure*}[t]
    \centerline{\includegraphics[width=0.9\textwidth]{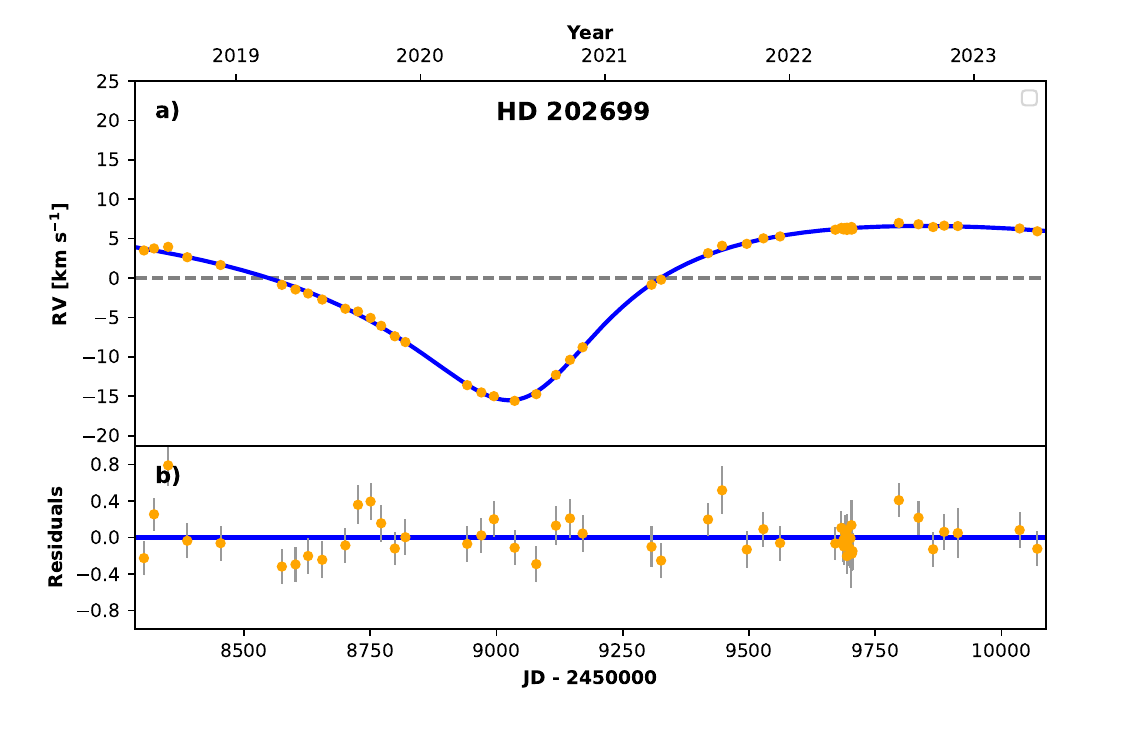}}
        \caption{RV curve of HD~202699: a) the complete measurements (circles) and RadVel fit (solid line),
        b) the residuals.\label{fig:HD202699}}
\end{figure*}

\begin{figure*}[t]
    \centerline{\includegraphics[width=0.9\textwidth]{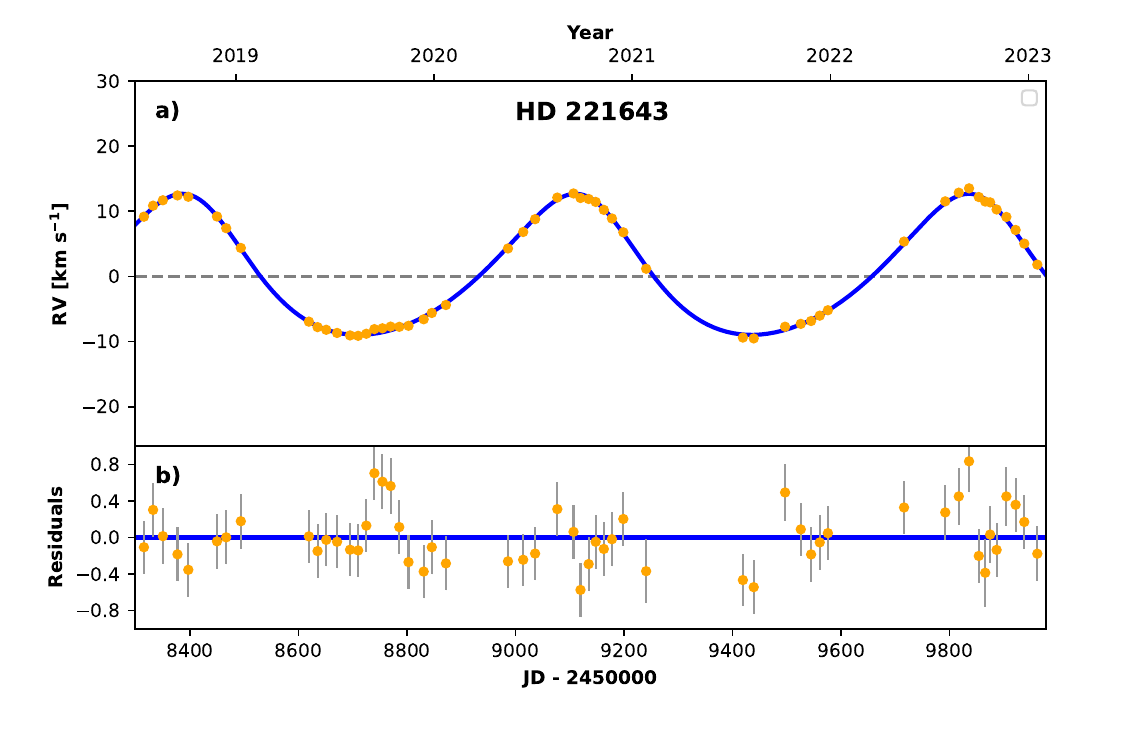}}
        \caption{RV curve of HD~221643: a) the complete measurements (circles) and RadVel fit (solid line),
        b) the residuals.\label{fig:HD221643}}
\end{figure*}

\end{document}